\documentclass[%
 preprint,
 amsmath,amssymb,
 aps,
superscriptaddress
]{revtex4-2}

\usepackage{graphicx}% Include figure files
\usepackage{dcolumn}% Align table columns on decimal point
\usepackage{bm}% bold math
\setcitestyle{super}

\usepackage{ulem}
\usepackage{upgreek}
\usepackage{color}
\usepackage{upgreek}
\usepackage[usenames,dvipsnames]{xcolor}
\newcommand{\Q}[1]{{{\textcolor{black}{#1}}}}

\begin{document}

%\preprint{APS/123-QED}

\title{Universal effective interactions \Q{of globular proteins} \\ close to liquid--liquid phase separation:\\ corresponding-states behavior\\ reflected in the structure factor}% Force line breaks with \\

	\author{Jan Hansen}
\affiliation{Heinrich Heine University, Condensed Matter Physics Laboratory, D\"{u}sseldorf, Germany}

	\author{Jannik N. Pedersen}
\affiliation{iNANO Interdisciplinary Nanoscience Center and Department of Chemistry, Aarhus University, DK-8000 Aarhus C, Denmark}

	\author{Jan Skov Pedersen}
\affiliation{iNANO Interdisciplinary Nanoscience Center and Department of Chemistry, Aarhus University, DK-8000 Aarhus C, Denmark}

	\author{Stefan U. Egelhaaf}
\affiliation{Heinrich Heine University, Condensed Matter Physics Laboratory, D\"{u}sseldorf, Germany}

	\author{Florian Platten}
	\email{florian.platten@hhu.de}
	\affiliation{Heinrich Heine University, Condensed Matter Physics Laboratory, D\"{u}sseldorf, Germany}
	\affiliation{Forschungszentrum J\"{u}lich, Institute of Biological Information Processing IBI-4, Biomacromolecular Systems and Processes, J\"{u}lich, Germany}

\date{\today}% It is always \today, today,

\begin{abstract}
Intermolecular interactions in protein solutions in general 
contain many contributions. If
short-range attractions dominate,  the state diagram exhibits liquid--liquid phase separation (LLPS) that is metastable with respect to crystallization.
In this case, the extended law of corresponding states (ELCS) suggests that thermodynamic properties are insensitive to details of the underlying interaction potential.
Using lysozyme solutions, we investigate
the applicability of the ELCS to the static structure factor and in how far effective colloidal interaction models can help to rationalize the phase behavior and interactions of protein solutions in the vicinity of the LLPS binodal. 
The (effective) structure factor has been determined by small-angle X-ray scattering (SAXS). It
can be described by Baxter's adhesive hard-sphere model, which implies a single fit parameter from which the normalized second virial coefficient $b_2$ is inferred and found to
quantitatively agree with previous results from static light scattering.
The $b_2$ values are independent of protein concentration, but systematically vary with temperature and solution composition, i.e. salt and additive content.
If plotted as a function of temperature normalized by the critical temperature, the values of $b_2$ follow a universal behaviour.
These findings validate the applicability of the ELCS to globular protein solutions and indicate that the ELCS can also be reflected in the structure factor.
\end{abstract}

\maketitle

\newpage
%\tableofcontents

\section{\label{sec:intro}Introduction}

Depending on the protein--protein interactions, especially if short-range attractions are dominant, protein molecules are prone to self-assemble into condensed states, like crystals, liquid-like droplets, and amorphous aggregates.\cite{Piazza2004,Vekilov2010,McManus2016}
Understanding the underlying interactions and their influence on these states and the protein phase behavior is beneficial for various fields of science and technology.
In medicine, the interactions that drive aggregation and, in particular,  liquid–liquid phase separation (LLPS) 
are relevant for intracellular organization and the regulation of biochemical reactions\cite{Brangwynne2015,Alberti2021} as well as the pathogenesis of severe diseases, like cataract and sickle-cell anemia.\cite{Pande2001,Galkin2002}
In structural biology, high-quality protein crystals are necessary for crystallographic structure determination and hence attempts to identify optimum protein crystallization conditions are highly desired.\cite{Chayen2008,Fusco2016}
In biopharmaceutics, the marginal stability of protein solutions against protein condensation poses a major challenge for formulation development,\cite{Saluja2008,Raut2016}
whereas the susceptibility to weak external stresses is exploited in food engineering to achieve functional product properties.\cite{Boire2019,Assenza2019}

However, protein--protein interactions are inherently complex due to the proteins' asymmetric molecular shape and heterogeneous surface properties reflected in, e.g., discrete charge patterns and the distribution of hydrophobic regions.\cite{Gunton2007}
The interaction potential is commonly assumed to comprise a repulsive contribution, owing to screened Coulomb interactions and steric hindrance, and an attractive part, including van der Waals and hydrophobic interactions.\cite{Israelachvili1992} 
In addition, hydrogen bonding, salt bridges and polymer-induced depletion interaction might also contribute.
The range and strength of the interactions do not only depend on the specific protein but are also modulated by other parameters, such as $p$H, temperature, ionic strength, solvent composition and protein concentration.
As a consequence, an adequate, quantitative description of protein--protein interactions and protein phase behavior remains challenging -- even on a coarse-grained level. Yet, in view of the importance in diverse fields and our incomplete understanding, attempts to further rationalize protein--protein interactions are highly desired.

Still, due to the size of the proteins, protein solutions can, on a coarse-grained level, be described by concepts developed in colloid science.
Although this neglects molecular details, it has proven helpful in understanding protein--protein interactions and protein phase behavior.\cite{Piazza2000,Stradner2020}
For example, the Derjaguin--Landau--Verwey--Overbeek (DLVO) theory has been used to describe the repulsive and attractive contributions to the interaction potential and to rationalize the dependence of inter-protein interactions on simple salts, different solvents and solvent mixtures or $p$H.\cite{Muschol1995,Poon2000,Pellicane2012,Goegelein2012,Hansen2021b}
Inspired by the DLVO theory, the structure factor of protein solutions, as probed by small-angle X-ray or neutron scattering, is sometimes modelled based on the sum of an attractive and a repulsive hard-core Yukawa potential each with its own range and interaction strength parameter.\cite{Tardieu1999,Liu2005c,Javid2007,Chinchalikar2013,Kundu2018}
Systems dominated by short-range attractions represent another example, particularly relevant for the present work.
In mixtures of colloids and small polymers, the polymers induce a short-range depletion attraction between the colloids and the gas--liquid coexistence can become metastable with respect to liquid--crystal coexistence.\cite{Lekkerkerker1992,Ilett1995,Poon2002}
Similar phase transitions have been observed for some square-well fluids\cite{Lomakin1996,Valadez-Perez2012} and patchy particle systems\cite{Sear1999,Liu2007,Goegelein2008,Kastelic2015}, and also
globular protein solutions can undergo metastable LLPS.\cite{Broide1996,Asherie1996,Muschol1997}
Accordingly, to describe scattering data in the vicinity of LLPS, various colloidal interaction models with short-range attractions have been employed, including square-well fluids, hard-core attractive Yukawa systems and adhesive hard spheres.\cite{Malfois1996,Zhang2007,Wolf2014,Braun2017,Braun2018}
Further attempts to rationalize the diversity of systems dominated by short-range attractions and the approaches to describe their interactions and phase behavior are desired.

It has been suggested\cite{Noro2000} that the thermodynamic properties of short-range attractive systems, including phase boundaries and the static structure factor,\cite{Zaccarelli2007,Ruiz-Franco2021} are insensitive to the details of the underlying interaction potential if the normalized second virial coefficient $b_2$ is used as a control parameter.
The second virial coefficient $B_2$ represents an integral measure of the interparticle interactions.\cite{Ben-Naim,McQuarrie} 
For a spherosymmetric potential $U(r)$ with center-to-center distance $r$, it reads
\begin{equation}\label{eq:b}
B_2 = 2 \pi \int_0^\infty \left( 1 - \exp{\left[-\frac{U(r)}{k_\text{B} T} \right]} \right) r^2 \text{d}r \, 
\end{equation}
with thermal energy $k_\text{B}T$. 
Its value is often reported as $b_2=B_2/(2\pi/3\,\sigma^3)$, where $B_2$ is normalized by the second virial coefficient of a corresponding hard-sphere system with diameter $\sigma$.  
As a consequence of the insensitivity to the details of the potential, the strength of the attraction as quantified by $b_2$ has been used as a predictor for gas--liquid and solid--liquid phase coexistence.\cite{Vliegenthart2000}
This so-called extended law of corresponding states (ELCS)\cite{Noro2000} has been validated for various model potentials.\cite{Valadez-Perez2012}
The ELCS has been shown to determine not only thermodynamic, but also local properties, such as cluster morphology.\cite{Valadez-Perez2018}
It is debated whether it also holds for the dynamics and non-equilibrium states, such as gels.\cite{Foffi2005,Lu2008,Zaccarelli2008,Gibaud2011}
Possible extensions of the ELCS to systems with competing interactions have been formulated and tested experimentally.\cite{Godfrin2014,Godfrin2018,Gruijthuijsen2018,Ruiz-Franco2021}
Moreover, its applicability to protein solutions with their complex interactions has been demonstrated for model proteins.
This includes studies on the metastable binodal of lysozyme solutions with different $p$H values, salt and additive concentrations\cite{Katsonis2006,Gibaud2011,Platten2015} as well as studies on the binodal, spinodal and osmotic compressibility of the lens protein $\gamma$B-crystallin at different H$_2$O/D$_2$O compositions\cite{Bucciarelli2016}.
\Q{If} the ELCS also applies to other thermodynamic properties of protein solutions, such as the static structure factor, \Q{they will be expected to show a corresponding-states behavior.} However, this has not been systematically explored \Q{so far}.

In the present work, 
the interactions \Q{of globular proteins} close to LLPS were examined by
small-angle X-ray scattering (SAXS). 
\Q{Here,} lysozyme in brine\Q{, a prime example for a system with short-range attractions, is used as a model system.
Moreover, lysozyme is commercially available in large amounts, allowing for quantitative and systematic studies.
As experimental state diagrams of lysozyme solutions are available,\cite{Goegelein2012,Platten2015,Platten2015b} links between the interactions and the state diagram can be explored based on SAXS experiments.}
The scattered X-ray intensity was determined for solutions with various protein, salt and additive concentrations and at different temperatures.
While the form factor was not affected by these changes, the structure factor at very small angles increased upon approaching LLPS, which was attained by increasing the protein concentration, decreasing temperature, or altering solution conditions. 
The structure factor contribution to the scattering was described by an adhesive hard-sphere model that depends only on one fitting parameter, namely $b_2$. 
The fit results indicate that, as expected from its thermodynamic definition,\cite{Ben-Naim} $b_2$ does not vary with protein concentration, but varies systematically with temperature and additive content.
For the solution conditions studied, the LLPS binodals show a universal behavior if the temperature axis is normalized by the critical temperature.
Accordingly, a universal temperature dependence of $b_2$ with respect to the critical temperature is observed.
Our results support the applicability of the ELCS to globular protein solutions and indicate its impact on the structure factor close to LLPS. 
%This work thus aims at a rationale for the interactions in protein solutions close to LLPS as provided by the ELCS.

\section{\label{sec:mm}Materials and Methods}

\subsection{Sample preparation}

Hen egg-white lysozyme powder (Roche Diagnostics, prod. no. 10837059001, purity $\geq 95~\%$), sodium chloride (NaCl; Fisher Chemicals, purity $\geq 99.8~\%$), guanidine hydrochloride (GuHCl; Sigma, prod. no. G4505, purity $\geq 99~\%$) and sodium acetate (NaAc; Merck, prod. no. 1.06268, p.a.) were used without further purification. 
Ultrapure water with a minimum resistivity of 18~M$\Omega$cm was used to prepare buffer and cosolvent stock solutions, which were filtered thoroughly (nylon membrane, pore size $0.2~\upmu\mathrm{m}$; VWR).
The protein powder was dissolved in a 50~mM NaAc buffer solution, which was adjusted to $p$H 4.5 by adding small amounts of hydrochloric acid. 
At $p$H 4.5 each lysozyme molecule carries approximately 11.4 positive net charges.\cite{Tanford1972}
Concentrated protein stock solutions were prepared by ultrafiltration, as described previously.\cite{Platten2015b}
Solution conditions are chosen to resemble those of our previous studies\cite{Goegelein2012,Platten2015b} to allow for a quantitative comparison. 
Samples were prepared by mixing appropriate amounts of buffer, protein and salt stock solutions 
and analyzed immediately after preparation. 
Sample preparation was performed at a temperature above the solution cloud-point to prevent immediate phase separation, typically at room temperature $(20\pm2)~^\circ\mathrm{C}$.
Few samples aggregated or crystallized during the measurements,\cite{Sedgwick2005,Hansen2016,Hentschel2021} as indicated by strongly increased scattering at very low angles; they were discarded from further analysis.
For selected conditions, samples were measured several times (up to six independently prepared and successfully measured samples) in order to check the reproducibility of our SAXS data and the validity of the resulting fit parameters.

\subsection{Small-angle X-ray scattering: Instrumentation}
Small-angle X-ray scattering (SAXS) was applied to determine the form factor and structure factor.
SAXS experiments were performed using the laboratory-based facilities at the Interdisciplinary Nanoscience Center (iNANO) at Aarhus University, Denmark,\cite{Behrens2014} as well as at the Center for Structural Studies at Heinrich Heine University D\"{u}sseldorf, Germany.
In Aarhus, a NanoSTAR SAXS camera (Bruker AXS) optimized for solution scattering\cite{Pedersen2004} with a home-built scatterless pinhole in front of the sample\cite{Lyngso2021} was used to measure the scattered intensity of protein and buffer solutions.
The solutions were filled in a thin flow-through glass capillary and thermostated using a Peltier element (Anton Paar) with a thermal stability of $0.1~^\circ\mathrm{C}$. 
In D\"{u}sseldorf, SAXS measurements on protein and buffer solutions were performed on a XENOCS 2.0 device with a Pilatus 3 300K detector.
The solutions were injected into a thin flow-through capillary cell mounted on a thermal stage (thermal stability $0.2~^\circ\mathrm{C}$).
Typical acquistion times of 10 and 5~min were used for dilute and concentrated solutions, respectively.
The data were background subtracted and converted to absolute scale using water (Aarhus)\cite{Pedersen2004} and glassy carbon (D\"{u}sseldorf) as standards.
The final intensity is displayed as a function of the magnitude of the scattering vector, $Q = (4\pi)/\lambda_0\sin(\theta)$, where the X-ray wavelength, $\lambda_0$, is $1.54~\text{\AA}$ and $2\,\theta$ is the angle between the incident and scattered X-rays and calibration was performed using silver behenate.

\subsection{Small-angle X-ray scattering: Data analysis}
The protein solutions are treated as monodisperse solutions of particles with small anisotropy and the particle positions are assumed to be independent of their orientations.
The absolute scattered intensity $I(Q)$ can then be described by the decoupling approximation:\cite{Kotlarchyk1983,Chen1986,Pedersen1997} 		
	\begin{equation}\label{eq:1}
	I(Q) = K\, c\, M\, P(Q) \, S_\text{eff}(Q) \, .
	\end{equation}
The $Q$ dependence of the scattered intensity is due to intra-particle and inter-particle interference effects quantified by the form factor $P(Q)$ and structure factor $S(Q)$, respectively. 
The form factor $P(Q)= \langle A^2(Q) \rangle_{\Omega}$ is obtained from the scattering amplitude $A(Q)$ averaged over particle orientations $\Omega$ (as denoted by brackets), and the effective structure factor\cite{Kotlarchyk1983,Chen1986,Naegele1996,Pedersen1997} reads
\begin{equation}
S_\text{eff}(Q)=1+ \frac{ \langle A(Q) \rangle^2_{\Omega} }{ \langle A^2(Q) \rangle_{\Omega} }  \left[ S(Q) - 1 \right] \, ,
\end{equation}
where $S(Q)$ is the structure factor of an effective one-component system.
The magnitude of the absolute scattered intensity depends on the particle (protein) mass concentration $c$, its molecular weight $M=14\,320~\text{g/mol}$, and the
contrast factor $K$, which is obtained by calibration and agrees with estimates\cite{Whitten2008}.

For very dilute systems, $S(Q) \approx 1$ and the $Q$ dependence of $I(Q)$ is dominated by the size, shape and structure of the individual particles via $P(Q)$.
On a coarse level,\cite{Pedersen1997} the form factor of lysozyme can be modelled as a prolate ellipsoid of revolution with minor and major axes as parameters. 
\Q{For the parameter range studied,} the form factor was found not to depend on the additive composition \Q{and temperature} (cf. Fig.~S1 of the Supplementary Material). It can be described by an ellipsoid using previously obtained parameters,\cite{Hansen2021b} namely a semi-minor axis $16.0~\mathrm{\AA}$ and an axial ratio $1.5$.

In concentrated solutions, the structure factor $S(Q)$ contains information on the spatial arrangement of the particles and thus reflects inter-particle interactions.
In our samples attractions dominate.
One of the simplest models to describe such systems is the adhesive hard-sphere (AHS) model proposed by Baxter.\cite{Baxter1968} 
An analytical approximation of the structure factor of adhesive hard spheres in the Percus-Yevick closure is available\cite{Regnaut1989,Menon1991,Menon1991b,Chen} and commonly used to model scattering data of short-range attractive systems\cite{Bergstrom1999,Eberle2012,Wolf2014,Matsarskaia2018}:

	\begin{equation} 
S(Q) = \frac{1}{1-C(Q)} \, ,  \label{eq:sq}
	\end{equation} 
with the Fourier transform of the direct correlation function multiplied by the number density\footnote{This corresponds to Eq.~(3) in Ref.~\citenum{Menon1991b}, whose left-hand side misses the multiplication by the number density.}
	\begin{equation*}
		\begin{split}
C(Q) = &  -\frac{24\,\phi}{(Q\,\sigma)^6}  \left\{ \vphantom{\frac{\phi\,\alpha}{2}}
 \alpha\,(Q\,\sigma)^3 \left( \sin{(Q\,\sigma)}-(Q\,\sigma)\,\cos{(Q\,\sigma)}\right) \right.
 \\ & \qquad \qquad
 + \beta\,(Q\,\sigma)^2\left[ 2\,(Q\,\sigma)\,\sin{(Q\,\sigma)}
		- \left((Q\,\sigma)^2-2\right)\,\cos{(Q\,\sigma)}-2\right]  \\ & \qquad \qquad \left.
 + \frac{\phi\,\alpha}{2} \left[ \left( 4\,(Q\,\sigma)^3-24\,(Q\,\sigma)\right)\sin{(Q\,\sigma)}-\left((Q\,\sigma)^4-12\,(Q\,\sigma)^2+24\right)\cos{(Q\,\sigma)}+24 \right] \right\} \\ 
&-\frac{2\,\phi^2\,\lambda^2}{(Q\,\sigma)^2}\left(1-\cos{(Q\,\sigma)}\right) + \frac{2\,\phi\,\lambda}{(Q\,\sigma)}\sin{(Q\,\sigma)}	\, . 
\end{split}
	\end{equation*} 
	The coefficients are given by:
	\begin{eqnarray*}
\alpha &=& \frac{\left(1+2\,\phi-\mu\right)^2}{\left(1-\phi\right)^4} \, , \\
\beta &=& - \frac{3\,\phi\left(2+\phi\right)^2-2\,\mu\left(1+7\,\phi+\phi^2\right)+\mu^2\left(2+\phi\right)}{2\left(1-\phi\right)^4} \,, \\
\mu& =& \lambda\,\phi\left(1-\phi\right) \, ,
	\end{eqnarray*}
	and $\lambda$ is the smaller root of the following equation:
\begin{equation*}
    \lambda \, \tau = \frac{\phi\,\lambda^2}{12}-\frac{\phi\,\lambda}{1-\phi}+\frac{1+\phi/2}{(1-\phi)^2} \, .
\end{equation*}  

The AHS structure factor depends on three parameters, namely the effective particle diameter $\sigma$, the stickiness $\tau$ and the particle volume fraction $\phi$.
The effective particle diameter $\sigma$ is identified with the diameter of a sphere that has the same volume as the ellipsoid determined by form-factor modelling in a dilute solution.
\Q{For consistency with previous studies\cite{Goegelein2012,Hansen2021b,Gibaud2011} and in agreement with densitometry\cite{Platten2015b},}
the volume fraction $\phi$ is obtained from $\phi = c / \rho_\text{P}$ with the protein concentration $c$ and (partial) mass density $\rho_\text{P} =1.351~\mathrm{g/cm}^3$.
Thus, only one fitting parameter, $\tau$, is involved, which is directly related to $b_2$:\cite{Chen}
\begin{equation}\label{eq:stick}
b_2 = 1-\frac{1}{4\tau} \,.
\end{equation}

The scattered intensity based on Eq.~(\ref{eq:1}) with a constant scattering background added is fitted to the measured scattered intensity using a least-square routine.
Since background subtraction is particularly delicate at very low $Q$, model fits are compared with experimental data for $Q \geq 0.025~\text{\AA}^{-1}$.
Further details on the data analysis have been given previously.\cite{Hansen2021b}

\section{\label{sec:res}Results and Discussion}

First, for conditions close but not extremely close to LLPS, the effects of protein concentration, temperature and additive concentration on the scattered X-ray intensity are examined. 
From the SAXS analysis, the corresponding effects on the underlying protein--protein  interactions are inferred, as quantified in terms of $b_2$.
Then, for the parameter range investigated, a universal LLPS phase boundary and a universal behavior of $b_2$ are observed if the temperature axis is normalized by the corresponding critical temperature $T_\text{c}$. 
Finally, the importance of these findings is discussed in the light of the ELCS.

Our investigation is intentionally limited to moderately concentrated solutions, $c\lesssim 100~\mathrm{mg/mL}$.
Furthermore, an extremely close proximity to the LLPS spinodal is avoided.
Thus, critical and off-critical scattering contributions \Q{as well as effects of the non-spherical protein shape on the structure factor} are expected to be small.
Hence, as validated below, analytical models for the structure factor can be reasonably employed to analyze the scattering data.
If instead \Q{protein solutions with} higher concentrations or temperatures very close to the spinodal line are considered, critical or off-critical contributions to the static structure factor will be expected, due to critical phenomena,\cite{Stanley1971,Dhont} and have previously been observed for the protein $\gamma_\text{B}$-crystallin.\cite{Bucciarelli2015,Bucciarelli2016}

\subsection{Effects of protein concentration and temperature on the interactions}

\begin{figure}
	\centering
	\includegraphics[width=0.95\textwidth]{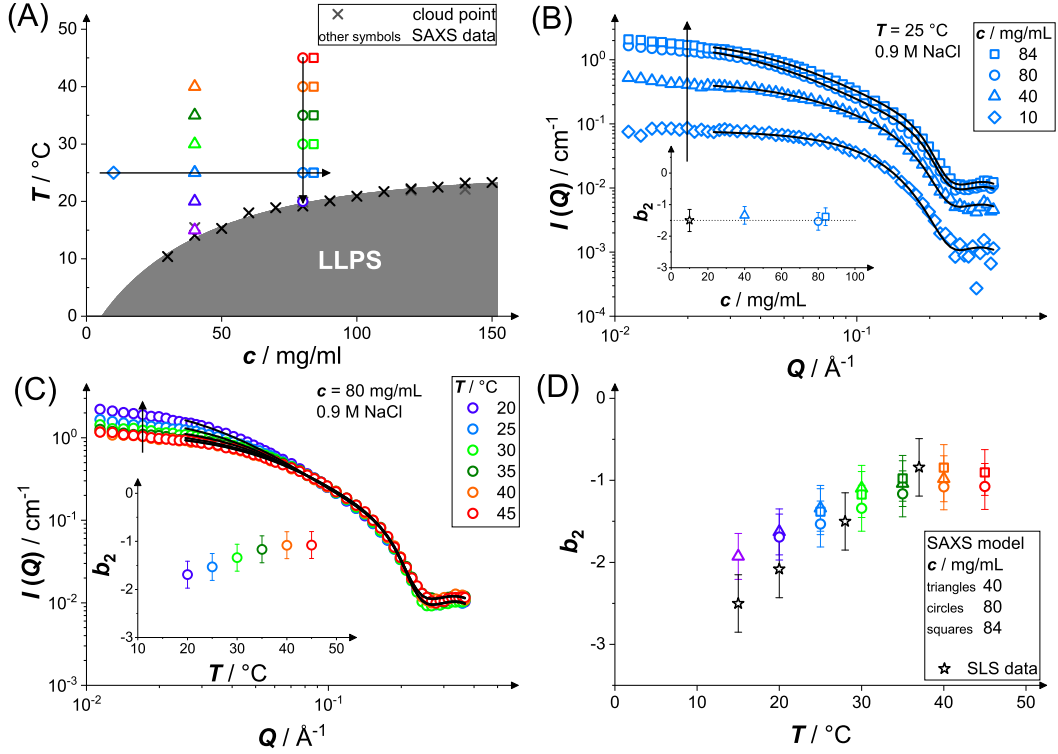}
	\caption{ 
			Protein solutions (lysozyme, $p$H 4.5, 0.9 M NaCl) close to metastable liquid--liquid phase separation (LLPS):
	(A) Temperature $T$ vs.\,\,protein concentration $c$ state diagram of protein solutions with the metastable LLPS boundary (cloud-point measurements\cite{Goegelein2012,Platten2015b} designated by crosses) and phase-separated region (grey-shaded area). 
	The other symbols mark the solution conditions explored by small-angle X-ray scattering (SAXS).
	The LLPS boundary can be approached by lowering $T$ or increasing $c$, as indicated by arrows.
	(B) Scattered X-ray intensity $I(Q)$ as a function of the magnitude of the scattering vector $Q$ of protein solutions with different $c$ (as indicated) at $T=25~^\circ\mathrm{C}$; data (symbols) and model fits (lines).
	The arrow indicates the approach to LLPS by increasing $c$.
	The inset shows the normalized second virial coefficient $b_2$, as retrieved from the fits to the SAXS data (colored symbols) and to light scattering data \cite{Goegelein2012} (black star displayed at a low $c$).
	(C) Scattered X-ray intensity $I(Q)$ as a function of the magnitude of the scattering vector $Q$ of protein solutions with  $c=80~\mathrm{mg/mL}$ at different $T$ (as indicated); data (symbols) and model fits (lines).
	The arrow indicates the approach to LLPS by lowering $T$.
	The inset shows the normalized second virial coefficient $b_2$, as retrieved from the fits (colored symbols).
	(D) Normalized second virial coefficient $b_2$ as a function of temperature $T$ for different $c$, as obtained from fits to SAXS data (colored symbols), and from light scattering\cite{Goegelein2012} (stars). 
}
\label{fig:1}
\end{figure}

%Figure~\ref{fig:1} shows data on the phase behavior (A),\cite{Goegelein2012,Platten2015b} the scattered X-ray intensity (B, C) and second virial coefficient (D) of protein solutions with various protein concentrations $c$ and temperatures $T$ as indicated. 
As in our previous works,\cite{Goegelein2012,Platten2015b,Hansen2016} the solutions contain a high salt concentration (0.9~M NaCl) that screens electrostatic repulsions and renders the interactions net attractive.
As a consequence, the solutions are metastable with respect to crystallization and undergo LLPS at low $T$.
Figure~\ref{fig:1}(A) illustrates the LLPS phase boundary (black crosses) as well as the solution conditions explored by SAXS (colored open symbols).
Arrows are used to indicate various ways toward the phase separated region (grey shaded area), either increasing $c$ or decreasing $T$.
(Note that for our solution conditions the LLPS spinodal is expected to be at least a few degrees below the binodal, as inferred from data for very similar solution conditions.\cite{Manno2003})

%The corresponding scattering data are displayed in Fig.~\ref{fig:1}(B) and (C), respectively. Additional scattering data for the remaining solution conditions (indicated but not marked by arrows) are shown in the Appendix (Fig.~\ref{fig:s2}(A,B)).

Figure~\ref{fig:1}(B) shows the scattered X-ray intensity (symbols) for protein solutions with different protein concentrations $c$ at $T=25~^\circ\mathrm{C}$, where clouding is expected to occur at much higher concentrations $c\approx 200~\mathrm{mg/mL}$\cite{Goegelein2012}.
With increasing $c$ the scattered intensity increases, as expected from Eq.~(\ref{eq:1}). 
The $Q$ dependence of the scattered intensity reflects both the form factor and the structure factor.
For the lowest concentration, $c=10~\mathrm{mg/mL}$, the data can be described by the form factor only, i.e., assuming $S(Q)=1$.
For higher $c$, variations in $I(Q)$ are due to changes of $S(Q)$ with $c$
as $P(Q)$ is expected to be independent of $c$. 
In particular, the $I(Q)$ at small and intermediate $Q$ reveal a more pronounced effect of the interactions with increasing $c$.

In order to quantify the net pair interactions, the model of Eq.~(\ref{eq:1}) is fitted to the experimental data using the stickiness $\tau$ as a free parameter (as well as the background).
The experimental data are quantitatively reproduced by the model fits (lines).
Then Eq.~(\ref{eq:stick}) is used to compute $b_2$.
The experimental uncertainty of $b_2$ is estimated to be $\pm 0.28$ based on the analysis of several independently prepared samples at the same condition.
The variation of $b_2$ with $c$ is displayed as an inset in Figure~\ref{fig:1}(B) (blue open symbols).
In addition, a value (black star) resulting from light scattering experiments\cite{Goegelein2012} is displayed at a low $c$.
In contrast to the present SAXS experiments, this literature result does not involve model fitting but is based on the $c$ dependence of $S(Q\to0)$.
The agreement between SAXS and light scattering data supports the appropriateness of our data analysis. 
Within the experimental uncertainties, $b_2$ is found to be constant and thus independent of $c$.
This is in line with thermodynamics;\cite{Ben-Naim,McQuarrie} $b_2$ is defined at infinite dilution and thus independent of $c$ and only dependent on $T$ (and on the particular solution environment under consideration). 

Figure~\ref{fig:1}(C) shows the scattered X-ray intensity for protein solutions at a fixed concentration $c=80~\mathrm{mg/mL}$ and various $T$, where clouding occurs at about $T=19.2~^\circ\mathrm{C}$.\cite{Goegelein2012}
Upon decreasing $T$, the scattered intensity at low $Q$ increases, reflecting an increased $S(Q\to0)$ and enhanced effect of the net attractions upon approaching LLPS.
The experimental data are quantitatively reproduced by the model fits.
Again, $b_2$ as obtained from the fits is displayed as an inset.
With decreasing $T$, $b_2$ becomes more negative, reflecting the changes of $S(Q)$, and thus quantifies the enhanced net attractions upon approaching phase separation.
Data with further concentrations are shown in the Supplementary Material (Fig.~S2(A, B)).

In Figure~\ref{fig:1}(D), $b_2$ data (colored symbols) for the solution conditions marked in Figure~\ref{fig:1}(A) are shown as a function of $T$.
As expected, at a fixed $T$, $b_2$ values obtained from fits to solutions with different $c$ are very similar.
Again, $b_2$ decreases when approaching LLPS by lowering $T$, reflecting increased attractions.
Moreover, the data agree with previous results\cite{Goegelein2012} from light scattering (black stars), though the latter show slightly lower $b_2$ values at low $T$.

\subsection{Effect of additive concentration on the interactions}

At a given protein concentration $c$, the temperature at which LLPS occurs can be regarded as a measure of the strength of the net attractions.\cite{Platten2016,Kastelic2015}
Accordingly, it depends on the solution conditions, such as the presence of salts\cite{Broide1996,Muschol1997} or solvents\cite{Goegelein2012,Hansen2021b} or the $p$H\cite{Taratuta1990}.
Here, guanidine hydrochloride (GuHCl), at molar concentrations a protein denaturant\cite{Greene1974}, is used at low concentrations to alter protein--protein interactions without affecting the protein size and shape (cf. Fig.~S1(A) of the Supplementary Material).
Guanidine can interact with proteins, e.g., via electrostatic interactions with charged and polar residues, hydrophobic interactions and hydrogen bonding,\cite{Liu2005,Mason2007,Raskar2019}
leading to reduced net attractions even at low concentrations as reflected in lowered solution cloud-points and crystallization boundaries.\cite{Platten2015b,Hansen2016}

\begin{figure}
	\centering
	\includegraphics[width=0.4\textwidth]{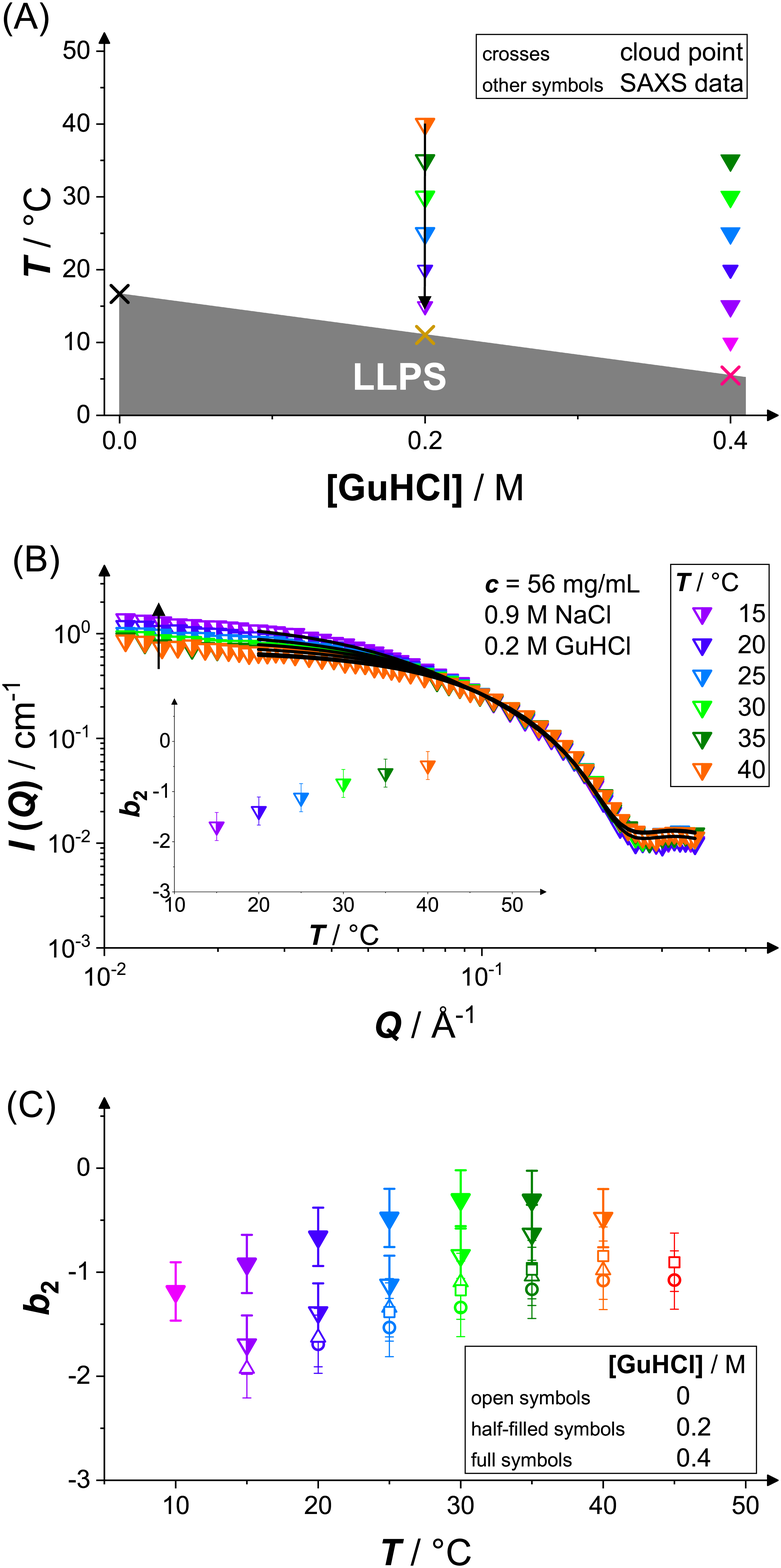}
	\caption{ 
	Protein solutions (lysozyme, $p$H 4.5, 0.9 M NaCl, $c=56~\mathrm{mg/mL}$) close to metastable liquid--liquid phase separation (LLPS):
	(A) Temperature $T$ vs.\,\,guanidine hydrochloride concentration $[\mathrm{GuHCl}]$ state diagram of protein solutions with the metastable LLPS boundary (cloud points, extrapolated from previous measurements,\cite{Platten2015b} designated by crosses) and phase-separated region (grey-shaded area). 
	The other symbols mark the solution conditions explored by small-angle X-ray scattering (SAXS).
	The LLPS boundary can be approached by lowering $T$, as indicated by an arrow.
	(B) Scattered X-ray intensity $I(Q)$ as a function of the magnitude of the scattering vector $Q$ for fixed protein and additive concentrations (as indicated) at different $T$; data (symbols) and model fits (lines).
	The arrow indicates the approach to LLPS by lowering $T$.
	The inset shows the normalized second virial coefficient $b_2$, as retrieved from the fits (colored symbols).
	(C) Normalized second virial coefficient $b_2$ as a function of temperature $T$ for different guanidine concentrations $[\mathrm{GuHCl}]$, as obtained from fits to SAXS data (colored symbols). 
}
\label{fig:2}
\end{figure}

For our solution conditions ($p$H $4.5$, $0.9~\mathrm{M}$ NaCl), the addition of GuHCl reduces the LLPS boundary by approximately $28~\mathrm{K/M}$, as illustrated for $c=56~\mathrm{mg/mL}$ in Figure~\ref{fig:2}(A).\cite{Platten2015b}
The solution conditions explored by SAXS are marked by colored half-filled and filled symbols.
Figure~\ref{fig:2}(B) shows the scattered X-ray intensity of a protein solution (symbols) in the presence of additional $0.2~\mathrm{M}$ GuHCl together with model fits (lines).
Again, upon approaching the LLPS by decreasing $T$, the low-$Q$ intensity increases as marked by a vertical arrow.
This indicates a stronger effect of the net attractions, as reflected in $b_2$ (displayed as an inset).
An additional data set in the presence of $0.4~\mathrm{M}$ GuHCl is provided in the Supplementary Material (Fig.~S2(C)).

Figure~\ref{fig:2}(C) shows $b_2$ data as a function of $T$.
The data for different $c$ in the absence of GuHCl shown in Figure~\ref{fig:1}(D) are replotted (small open symbols).
Data in the presence of various amounts of GuHCl (half-filled and filled symbols) are shown in addition.
For fixed guanidine concentration, $b_2$ decreases when $T$ is lowered, again reflecting enhanced attractions while approaching LLPS.
At a fixed $T$, $b_2$ increases with the concentration of GuHCl, indicating reduced net attractions consistent with the decreased proximity to the LLPS phase boundary (Fig.~\ref{fig:2}(A)).

\subsection{Universal phase boundary and effective interactions}

\begin{figure}
	\centering
	\includegraphics[width=0.5\textwidth]{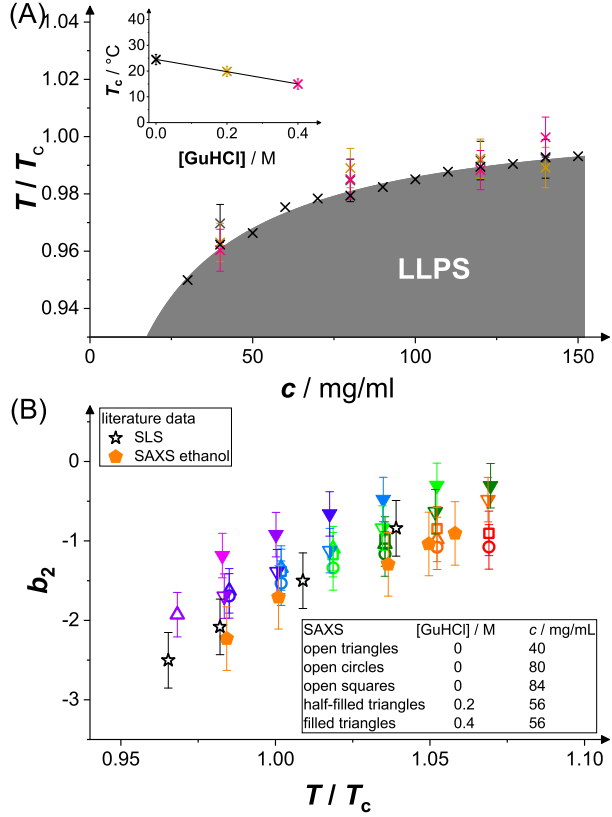}
	\caption{ 
	Protein solutions (lysozyme, $p$H 4.5, 0.9 M NaCl) with different additive composition close to metastable liquid--liquid phase separation (LLPS):
	(A) Temperature $T$ normalized by the respective critical temperature $T_\mathrm{c}$ (as provided in the inset) vs.\,\,protein concentration $c$ state diagram of protein solutions with the metastable LLPS boundary (cloud-point measurements\cite{Goegelein2012,Platten2015b} designated by crosses) and phase-separated region (grey-shaded area). 
	(B) Normalized second virial coefficient $b_2$ as a function of the normalized temperature $T/T_\mathrm{c}$, as obtained from the analysis of SAXS experiments (present data and literature results\cite{Hansen2021b}) and from light scattering\cite{Goegelein2012}. 
}
\label{fig:3}
\end{figure}

According to the two previous sections, the interaction strength quantified by $b_2$ is independent of $c$ (Fig.~\ref{fig:1}(B), inset), but systematically varies with $T$ (Fig.~\ref{fig:1}(D)) and solution composition (Fig.~\ref{fig:2}(C)).
In order to further rationalize our findings, the extended law of corresponding states is invoked:
Thermodynamic properties, such as vapor pressure or liquid--vapor coexistence, follow a master curve if temperature, density and pressure are normalized with respect to their values at the critical point.\cite{Pitzer1939,Guggenheim1945}
In analogy, for protein solutions, the critical LLPS temperature $T_\text{c}$ can be considered as an integral measure of the net attractions that are present for a particular solution condition.\cite{Platten2015,Platten2016}
If LLPS coexistence curves for different solution conditions are normalized by their $T_\text{c}$ values (and the repulsive interactions are alike), the normalized LLPS curves tend to follow a master curve.\cite{Katsonis2006,Platten2015,Bucciarelli2016} 
Following this approach, we estimate
$T_\text{c}$ assuming a critical exponent for binary demixing from renormalization-group theory,\cite{Platten2015,Hansen2016,Hansen2021b} \Q{as detailed in the Supplementary Material.}
The normalized LLPS coexistence curves are displayed in Figure~\ref{fig:3}(A) and the $T_\text{c}$ values are shown as an inset.
Similar to the decrease of the LLPS temperatures (Fig.~\ref{fig:2}(A)), $T_\text{c}$ decreases upon addition of GuHCl.
As a consequence, the net interactions for the different solution conditions are expected to depend only on the relative temperature $T/T_\text{c}$.
Since the repulsive interactions are weak and similar, this implies that the attractive interactions only depend on the relative temperature $T/T_\mathrm{c}$.

Figure~\ref{fig:3}(B) shows the $b_2$ data from Figure~\ref{fig:2}(C) as a function of the normalized temperature $T/T_\mathrm{c}$. 
Within experimental uncertainty, the data tend to follow a universal behavior.
In addition, light scattering data on lysozyme in brine\cite{Goegelein2012} and SAXS data on lysozyme in the presence of NaCl and water--ethanol mixtures\cite{Hansen2021b} are included.
These literature data and the new results agree and thus also follow the universal behavior.
These results imply that, if critical and off-critical scattering are negligibly small, the orientationally-averaged structure factor close to LLPS is controlled by three parameters, the effective particle diameter, the protein concentration (or volume fraction) and the relative temperature $T/T_\mathrm{c}$ (or, equivalently, the second virial coefficient $b_2$). If the AHS model is applicable, then these parameters are sufficient to predict the (effective) structure factor based on Eq. (\ref{eq:sq}) and (\ref{eq:stick}), supporting and broadening the applicability of the extended law of corresponding states to globular protein solutions.
\Q{Experimental systems and theoretical models dominated by short-range attractions do not only include protein solutions and adhesive hard spheres, respectively, but also nanoparticle dispersions and colloid-polymer mixtures\cite{Lekkerkerker1992,Ilett1995,Poon2002} as well as patchy particle systems\cite{Sear1999,Liu2007,Goegelein2008,Kastelic2015}. }
One could speculate that, not only in the limit $Q\to0$, but also for small $Q$, the structure factors of various \Q{such} systems close, but not extremely close to LLPS are very similar to those of the AHS model; this, however, needs to be tested by systematic theoretical work.

\section{\label{sec:con}Conclusion}
  
The scattered X-ray intensity of protein solutions in the vicinity of liquid--liquid phase separation was determined for different protein concentrations, temperatures and additive concentrations.
The structure factor was modelled based on the adhesive hard-sphere model.
A fit to the data yielded the normalized second virial coefficient $b_2$.  
It was found to be independent of protein concentration, but to vary with temperature and additive concentration.
The results agree with previous findings, in particular with model-independent results from light scattering.
If the temperature is normalized by the LLPS critical temperature $T_\text{c}$, the second virial coefficient $b_2$ follows a universal dependence on the normalized temperature $T/T_\text{c}$.
This suggests that also the protein--protein interactions and hence the (effective) structure factor exhibit a universal dependence on $T/T_\text{c}$.
Our results thus support and broaden the applicability of the extended law of corresponding states (ELCS), here tested for globular protein solutions.

\section*{Supplementary Material}

\Q{See Supplementary Material for additional SAXS data and for details of the estimation of the critical temperature.}

\section*{Acknowledgements}
We thank Gerhard N\"{a}gele (J\"{u}lich, Germany) and Ram\'on Casta\~neda-Priego (Leon, Mexico) for very helpful discussions.
F.P. acknowledges financial support by the Strategic Research Fund of the Heinrich Heine University (F 2016/1054-5) and the German Research Foundation (PL 869/2-1).
We thank the Center for Structural Studies (CSS) for access to the SAXS instrument.
CSS is funded by the Deutsche Forschungsgemeinschaft (DFG Grant numbers 417919780 and INST 208/761-1 FUGG).

\section*{AUTHOR DECLARATIONS}

The authors declare no conflicts of interest.

\section*{DATA AVAILABILITY}
  
The data that supports the findings of this study are available within the article.

\normalem
\bibliography{lorena}  
  
\end{document}